\documentclass[sigconf,authorversion,nonacm]{acmart}

\AtBeginDocument{%
  \providecommand\BibTeX{{%
    \normalfont B\kern-0.5em{\scshape i\kern-0.25em b}\kern-0.8em\TeX}}}

\copyrightyear{2021}
\acmYear{2021}
\setcopyright{rightsretained}
\acmConference[CHI PLAY '21]{Extended Abstracts of the 2021 Annual Symposium on Computer-Human Interaction in Play}{October 18--21, 2021}{Virtual Event, Austria}
\acmBooktitle{Extended Abstracts of the 2021 Annual Symposium on Computer-Human Interaction in Play (CHI PLAY '21), October 18--21, 2021, Virtual Event, Austria}
\acmDOI{10.1145/3450337.3483484}
\acmISBN{978-1-4503-8356-1/21/10}

\begin{document}

\title[Dota 2: Better Frame Rates or Better Visuals?]{Better Frame Rates or Better Visuals? \\ An Early Report of Esports Player Practice in Dota 2}

\author{Arjun Madhusudan}
\affiliation{%
  \institution{North Carolina State University}
  \city{Raleigh}
  \state{North Carolina}
  \country{USA}
  }
\email{amadhus2@ncsu.edu}

\author{Benjamin Watson}
\affiliation{%
  \institution{North Carolina State University}
  \city{Raleigh}
  \state{North Carolina}
  \country{USA}
  }
\email{bwatson@ncsu.edu}

\begin{abstract}
  Esports athletes often reduce visual quality to improve latency and frame rate, and increase their in-game performance. Little research has examined the effects of this visuo-spatial tradeoff on performance, but we could find no work studying how players manage this tradeoff in practice. This paper is an initial examination of this question in the game Dota 2. First, we gather the game configuration data of Dota 2 players in a small survey. We learn that players do limit visual detail, particularly by turning off VSYNC, which removes rendering/display synchronization delay but permits visual “tearing”. Second, we survey the intent of those same players with a few subjective questions. Player intent matches configuration practice. While our sampling of Dota 2 players may not be representative, our survey does reveal suggestive trends that lay the groundwork for future, more rigorous and larger surveys. Such surveys can help new players adapt to the game more quickly, encourage researchers to investigate the relative importance of temporal and visual detail, and justify design effort by developers in "low visual" game configurations.
\end{abstract}



\maketitle

\section{Introduction}
With high-speed internet, online gaming has become increasingly common, giving rise to a professional esports industry that rivals traditional sports. Like those traditional sports, esports has begun driving down the price of its technologies, making graphics, display and input equipment available at low prices for casual gamers. 

As in most sports, both elite and casual esports athletes seek an edge. This can mean improved temporal detail, in the form of frame rate and latency; or better visual detail, such as additional rendering or animation. Prevailing opinion in the esports community prefers temporal over visual detail. For example, many advocate turning off VSYNC, which increases frame rate and reduces latency by removing synchronization delay between the graphics processing unit (GPU) and the display (e.g. \cite{pcgamehaven}). The tradeoff introduced with this improvement is \textit{tearing}, as two or more frames --- each showing a different moment in time --- appear on the display. 

The combined effect of temporal and visual detail on performance in games has received limited study, but we are not aware of any research on how players actually make this tradeoff in practice. Given the growing number of esports athletes wrestling with this same question, we performed a pilot survey of how esports athletes manage the visuo-temporal tradeoff in their own gameplay.

\section{Visual and Temporal Detail}
In computer games, configurable visual detail typically includes features that players can see, but have limited effect on gameplay. Instead, they are primarily a way of making the game more pleasing, emphasizing the fantasy element of the game-world. Visual detail typically includes options like higher-resolution texture, ambient occlusion, more accurate lighting, special animation, and even UI elements such as an animated/live heads-up display (HUD).
   
Options for configuring temporal detail are more limited, and can include VSYNC (de)activation and frame rate caps. Instead, because visual and temporal detail are in tradeoff, players often configure temporal detail by manipulating visual detail. For example, a player might lower texture detail to increase frame rate and reduce latency. Both the research and the esports communities believe that temporal detail  plays a major role in gaming performance.

Esports players can configure settings and the visuo-spatial tradeoff as they like, but can become quite confused while doing so: each of the dozens of relevant settings is hard to understand and difficult to find among hundreds of other irrelevant settings. 

\section{Related work}

Improved temporal detail has long been found to improve human performance in general \cite{chen2007review}. In games, reducing latency and increasing refresh rates often translates into winning. For example, even reducing latencies from 50 to 25 milliseconds continues to improve player performance in CS:GO (Counterstrike Global Offensive) \cite{10.1145/3411764.3445245}. 240 Hz monitors and adaptive sync have proven to be useful to players \cite{kim2019esports,10.1145/3320286}, even when they could not tell when they were using them. Below 60 frames per second (FPS), frame rate often affects game performance more than latency \cite{10.1145/2559206.2581214}. However, over 60 FPS, the effect of frame rate on performance lessens \cite{10.1145/3355088.3365170}. 

Research on how visual detail affects performance is limited. Claypool et al. \cite{claypool,claypool2009perspectives} are among the few who examined a possible relationship, and could not confirm it. They did however find a relationship to player experience, and indeed visual elements play an important role in the gaming economy. Free to play games like Dota 2 thrive on monetization of virtual goods. These include non-functional animations, effects, and in-game cosmetics such as player skins and avatars \cite{10.1145/3025453.3025893}. Their quality is highly influenced by detail settings, giving players an incentive to use higher visual detail.  

Even less research addresses the combined effect of temporal and visual detail on gaming performance, with Claypool et al. \cite{claypool,claypool2009perspectives} finding that resolution did not interact with frame rate's effects.

While most research on the visuo-temporal tradeoff in gaming has examined its effect on player performance, this paper focuses on how players choose to make this tradeoff in their own gaming.

\section{Data collection}
We chose to study the visuo-temporal detail tradeoff in the context of Dota 2, one of the most popular esports games. Dota pits networked opponents against one another in a 3D environment, with the goal of taking the opponent's base. The quick pace of the game provides an advantage to the first mover, and rewards players with quick reflexes and faster computer systems. The game has over one million active players, with its largest annual tournament ``The International'' awarding a prize pool of \$40,000,000  in 2020. Dota 2 also adopts a multiplayer online battle arena (MOBA) style of play, which is quite common among the most significant esports games. 

To collect information regarding the in-game configurations of Dota 2 players, we recruited from Reddit --- particularly r/dota2\cite{rdota2} and its child subreddits. We asked respondents to submit their Dota 2 configuration files, each of which describes over 450 in-game settings. We requested that players obtain these files from the same system they used to play Dota 2. The files are hard to understand, but automatically generated based on player menu choices, and placed in a subdirectory that is difficult to locate. This gives us some confidence that the files we received were unaltered. We also asked respondents to answer a three-question subjective survey:

\begin{itemize}
    \item {Have you played Dota 2 competitively (on any scale)?}
    \item {Which setting would you reduce to increase computer performance (reduce lag/system crashes etc)?}
    \item {Which setting would you reduce to increase your own performance (details that might distract you or make it harder to see while playing etc)?}
\end{itemize}

To the first question, respondents could reply ``no'' or ``yes,'' followed by a short description of the tournament. To answer the second and third questions, respondents ordered four options: ``shadows,'' ``textures,'' ``screen size,'' or ``FPS''.

\subsection{Participants}

Our survey was open to any gamer who used Windows to play Dota 2, and who actively played Dota 2 on any level, whether casual or professional. No personal information was collected, and no compensation was provided. We received 43 responses, with one respondent submitting an invalid configuration file.

We do not claim that our sample of 43 players is representative of the Dota 2 player base, which numbers in the millions \cite{dota2numbers}, and is internationally popular \cite{dota2countries}. Nevertheless, our preliminary survey does suggest trends and raise questions that might be followed up in more rigorous future surveys.

\section{Results}
We begin by studying the configuration files we collected, and then examine responses to our three-question subjective survey.

\subsection{Configuration results}

All the configuration files parsed correctly. From the over 450 settings in each configuration file, we focused on 15 that affected visual or temporal game detail, and which players could easily change through the in-game configuration screen (see Table \ref{tab:config_data}). 


\begin{table}[!h]
\centering
\begin{minipage}[b]{0.5\textwidth}
  \centering
    \caption{The 15 relevant configuration settings, detailed in Appendix \ref{app:settings_config}. Columns from left to right show the setting identifier, a sparkline of that setting's participant data, and the proportion of that setting's data that limited visual detail.}
    \label{tab:config_data}
    \includegraphics[width=\linewidth]{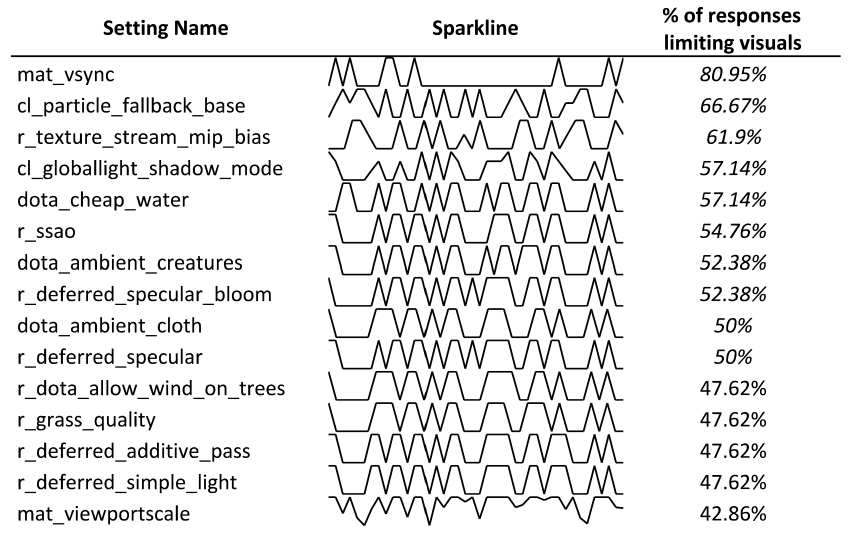}
\end{minipage}%
\end{table}


Definitions of some key settings include:
\begin{itemize}
    \item {\textit{useadvanced}: Switches between simple and complex setting controls. 0 manages settings with preset groups, while 1 allows the player to customize each setting individually.}
    \item {\textit{mat\_vsync}: Indicates whether the player has VSYNC on or off. When VSYNC is on, GPU frames and display refreshes are synchronized, reducing frame rate but limiting tearing. When it is off, frame rate increases but tearing is visible.}
    \item { \textit{r\_texture\_stream\_mip\_bias}: Changes texture resolution in the mipmap. 0 uses coarse textures, 1 medium, and 2 fine. We considered values 0 and 1 as reduced visual detail.}
    \item { \textit{mat\_viewportscale}: Scaling factor for the viewport. If less than 1, it limits visual detail by rendering fewer pixels.}
    \item { \textit{cl\_particle\_fallback\_base}: Controls level of detail in special effects such as glow, particles, and smoke, with 0 being low, 2 being medium and 4 being high. We considered values 0 and 2 as limiting visual detail, while 4 did not.}
\end{itemize} 

While we could not locate one definitive resource that clearly defines all Dota 2 settings, most can be found in online Dota 2 forums \cite{redditpost, steamcommunity}. With a few exceptions  (\textit{texture\_stream\_mip\_bias}, \textit{mat\_viewportscale}, \textit{cl\_globallight\_shadow\_mode}, and  \textit{particle\_ fallback\_base}), all settings had two possible values. 

Table \ref{tab:config_data} contains setting percentages and sparklines. The only settings over 60\% limited visuals: \textit{vsync}, \textit{texture\_stream\_mip\_bias} and  \textit{particle\_fallback\_base}. In particular, \textit{vsync} was turned off by over 80\% of participants, improving temporal detail (frame rate) while harming visual detail (tearing). Sparklines traditionally show variation vs. time, we use them to visualize settings vs players. Similar sparklines reveal that most participants configured most settings in much the same way. Many settings are configured automatically by Dota 2 or graphics drivers. To get a better sense of which settings were then altered directly by our participants, we considered only those players who used advanced mode (\textit{useadvanced}) to customize individual Dota 2 settings. \textit{useadvanced} is never enabled on installation, meaning a participant must act to enable it, making it more likely that they will determine other settings manually.


\begin{table}[!ht]
\centering
\begin{minipage}[b]{.5\textwidth}
  \centering
    \caption{The same 15 configuration settings as in Table \ref{tab:config_data}, but only for participants who are likely more familiar with them, and chose to use advanced settings.}
    \label{tab:config_data_advanced}
    \includegraphics[width=\linewidth]{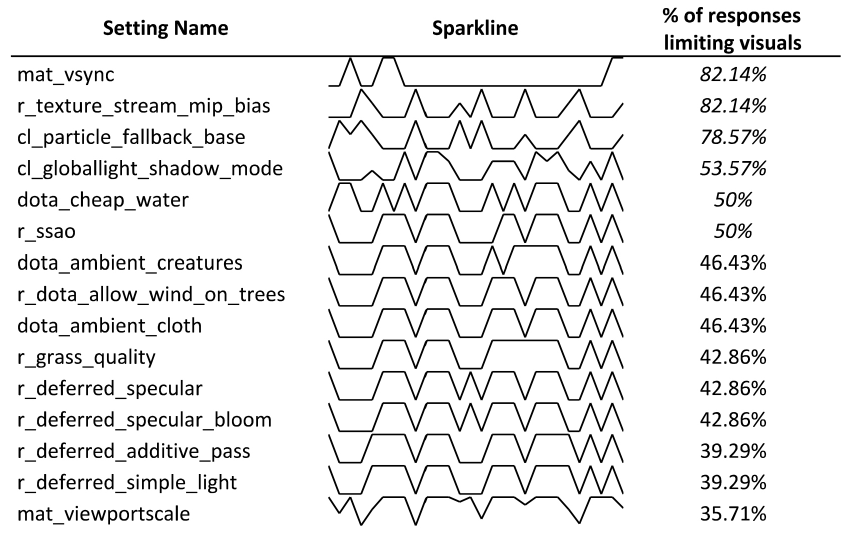}
\end{minipage}%
\end{table}

Table \ref{tab:config_data_advanced} shows how the same 15 settings were configured by participants that enabled \textit{useadvanced}. The settings show strengthening trends: now both \textit{vsync} and \textit{texture\_stream\_mip\_bias} limit visuals for more than 80\% of players, and \textit{particle\_fallback\_base} nearly does the same. In addition, new trends limiting temporal detail emerge: \textit{viewportscale}, \textit{deferred\_additive\_pass} and \textit{simple\_light} all limit temporal detail (and improve visual detail) for more than 60\% of players. Participants’ settings were also slightly less consistent, with the \textit{cheap\_water} setting now more different from following settings than among the full 42 participants.

\subsection{Survey results} \label{surveyres}

Our participants represent competitive and casual gamers in roughly equal proportion. Of the 43 players who responded, 21 players (48.8\%) played the game competitively, meaning they took part in tournaments on various levels such as school, nation or online competitive platforms such as FaceIt.

\begin{figure}[htb]
\centering
\begin{minipage}{.48\textwidth}
  \centering
  \includegraphics[width=\linewidth]{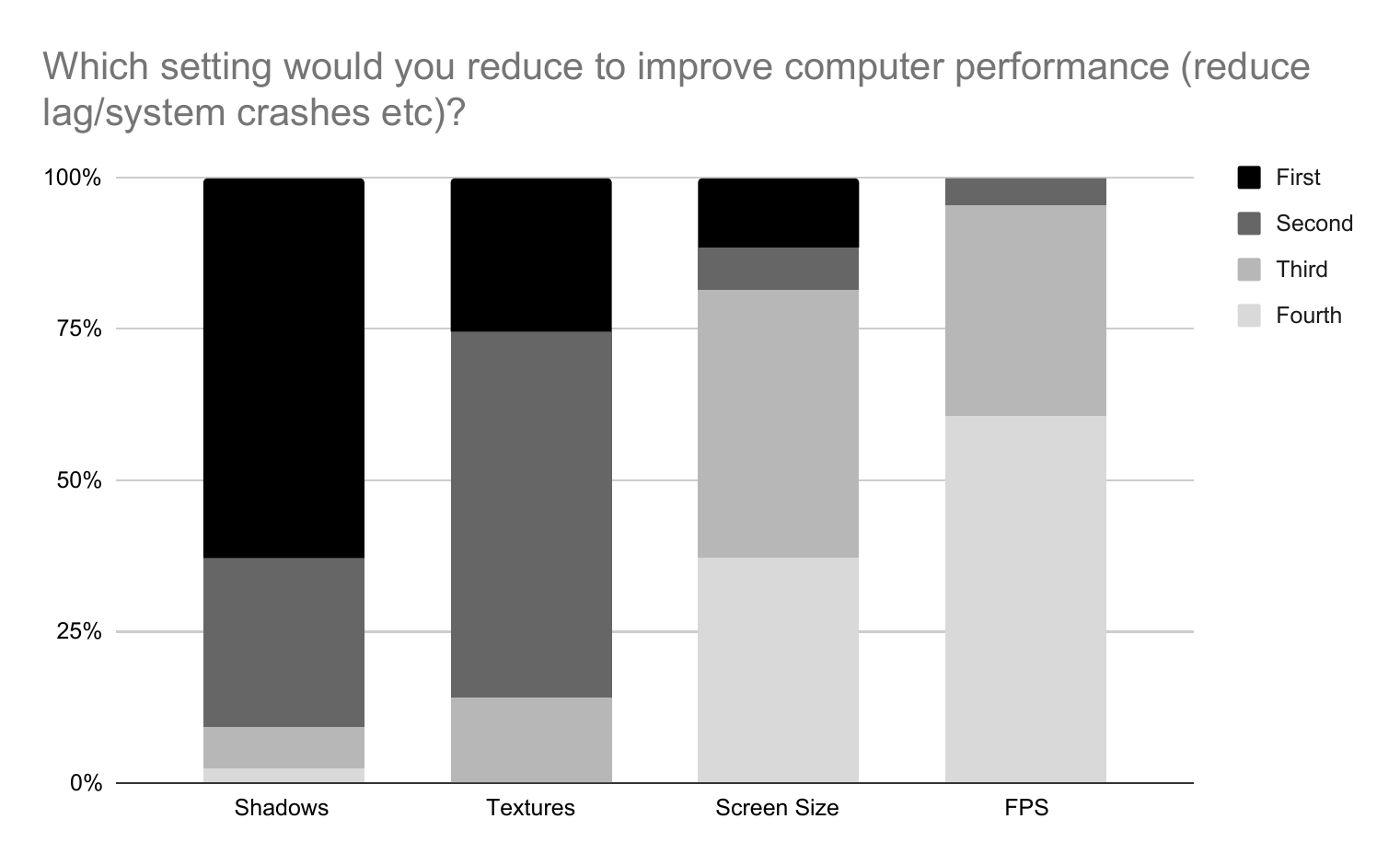}
  \captionof{figure}{Participants' orderings of methods for improving computer performance.}
  \label{fig:computer_perf}
\end{minipage}%
\end{figure}

Participants' answers to the second (“improve computer performance”) and third (“improve your own performance”) questions were similar. Figure \ref{fig:computer_perf} shows participants’ answers to the second question. 27 players (62.7\%) would reduce shadows first to improve system performance, while 26 players (60.5\%) would limit FPS last. 35 players (81.4\%) would not reduce screen size first or second.

\begin{figure}[htb]
\centering
\begin{minipage}{.48\textwidth}
  \centering
  \includegraphics[width=\linewidth]{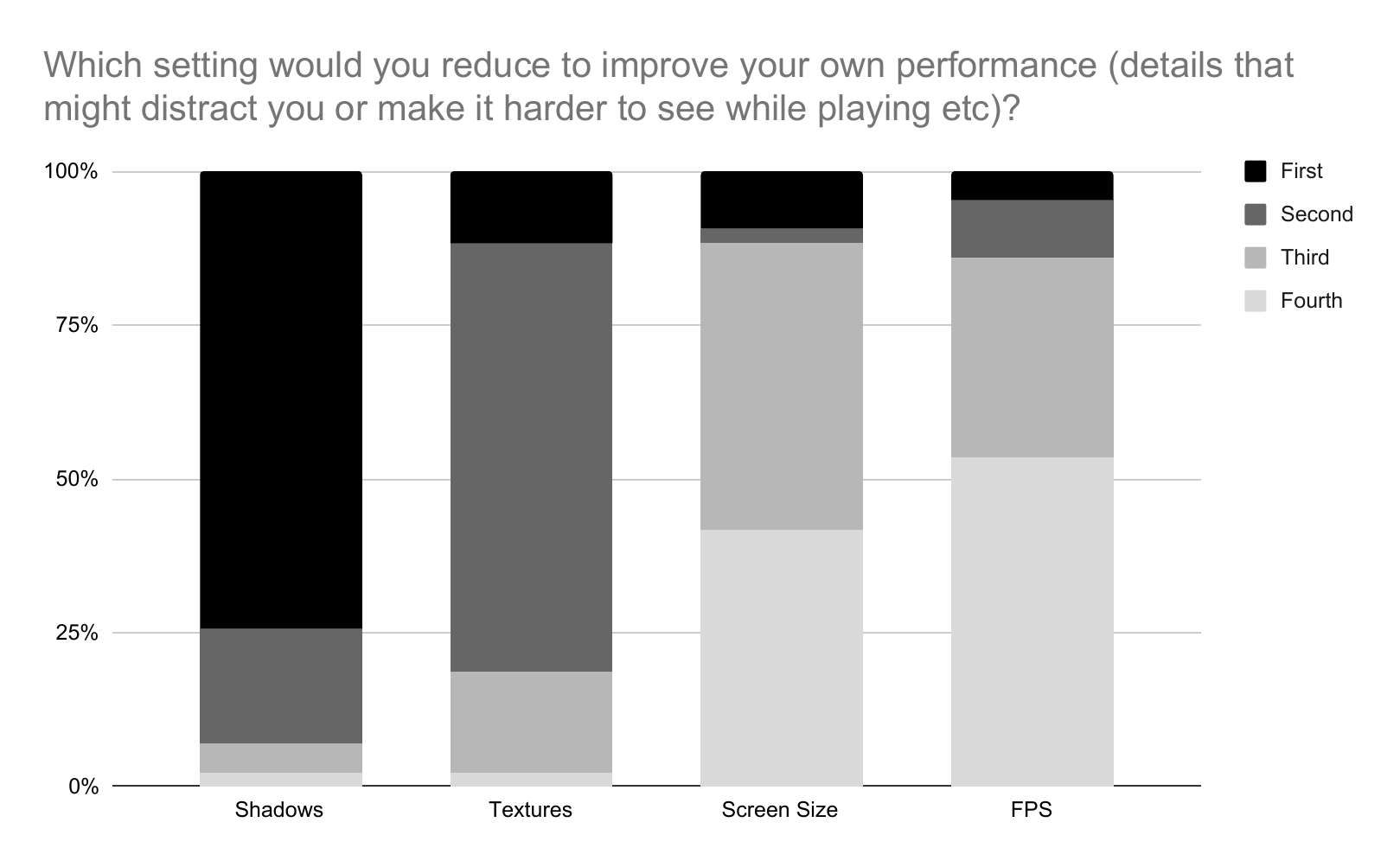}
  \captionof{figure}{Participants' orderings of methods for improving their gaming performance.}
  \label{fig:gaming_perf}
\end{minipage}
\end{figure}

Figure \ref{fig:gaming_perf} shows participants’ answers to the third question. 32 players (74.4\%) said they would first reduce shadows to improve their own performance, while 23 players (53.3\%) agreed that limiting FPS would be their last option. 38 players (88.4\%) did not place reducing screen size first or second on their list.

\section{Discussion}
We begin our discussion of these results by reviewing some of the many reasons for caution in interpreting them. We then move on to possible explanations and implications.

\subsection{Limitations}
There are many features in this work that limit its generality, and suggest possible future work. We discuss a few of these here.

First, our 43 respondents are not a representative sample of Dota 2 players, which numbers in the millions. Despite this large population, survey methodology should allow future researchers to obtain a representative sample with responses numbering in the low thousands. As groundwork for such an effort, we need a good sense of Dota 2 demographics across (perhaps) nationality, game familiarity, age and gender --- an ``esports census''. 

Next, we cannot be certain whether settings result from conscious player choice, or are automated system defaults. Future studies might ask respondents to identify those settings they made themselves. Information about respondent hardware configurations (e.g. CPU, GPU) might put settings in an informative context. Furthermore, having the players take us through their settings in a think-aloud fashion would provide valuable information.

In addition, even when players make conscious choices, we cannot be certain that they understood their implications. As this paper begins to make clear, the range of settings and possible interactions they might have with one another, hardware, and gameplay can be overwhelming. For example, a player might choose a setting because they believe it limits a visual effect and improves frame rate, but they are quite mistaken. Future work might query player understanding of Dota 2 settings. 

Finally, Dota 2 is just one of several dominant esports games, and one of hundreds of recently significant titles in the broader gaming industry. These games can vary significantly in genre (e.g. battle area vs. first-person shooter), task and more. The trends we find here may not be found in other games.

\subsection{Possible explanations}
\label{explanations}
In our three-question survey, the similarity between replies to the ``improve computer performance'' and ``improve your own performance'' questions suggests that participants associated gaming performance with computer performance, not with visual quality: the faster their system ran, the better they would play. We can see these survey results reflected in respondents configuration settings as well: textures and shadows are usually limited. However, in both the survey and configuration settings, rendered screen size (\textit{viewportscale}) was usually not reduced, despite the improvements in temporal detail that would result. It may be that this particular type of visual detail --- the number of pixels rendered, which has a much more global visual impact than other types, such as texture or shadows --- is especially valuable to players.

The strongest trend in our configuration results is the widespread deactivation of VSYNC, which improves temporal detail as measured by frame rates and latencies, but harms visual detail by introducing a tear into displayed imagery between frames representing different moments in time. Players seem quite willing to sacrifice temporal coherence for more up-to-date information. To our knowledge, these are the first (admittedly preliminary) data describing this preference.

The remaining strong trends in our configuration settings data indicate that players regularly limit visual effects such as texture resolution and animated particle complexity to improve temporal detail. However, weaker trends among better informed and more proactive players (using advanced settings) show that players sometimes improve visual effects (e.g. by maintaining screen size or using an additive rendering pass). As we speculated with our screen size discussion above, it may be that certain types of visual detail are particularly important. Alternatively, they may have only a minor impact on temporal detail and/or player performance.

In both our broader and advanced configuration data, we observed a great deal of within-player consistency, that is, players consistently did/did not limit visual detail. This might reflect individual player goals, or alternatively, consistent defaults made by the player’s system.

\subsection{Implications}
We consider the implications of our pilot esports configuration survey for for players, designers and researchers.

Players should prioritize improvements in temporal detail when configuring their games. Prior research on temporal detail's effects on gaming performance shows that better frame rates and latencies improve gaming performance, and indeed in our survey the configurations of most players prioritize improvements in frame rates. Interestingly however, not all reductions in visual detail are justified; players should carefully evaluate each setting's effect on gameplay.

Given the importance of temporal detail for game performance, designers may wish to offer additional settings with good ``temporal/visual ratios'', that is, that maximize the temporal benefits and minimize the visual penalties for gaming performance. Additionally, the obtuse and tangled collection of settings can easily overwhelm both players and researchers: please simplify and explain! 

In addition to scaling and generalizing this work, researchers might investigate why some visual tradeoffs were made (e.g. \textit{r\_ texture\_stream\_mip\_bias}) and others were not (e.g. \textit{mat\_viewportscale}). Could it be related to gaming experience, as opposed to performance? And finally, the complex web of detail settings is a constant source of frustration and confusion for users. Solving this problem could be important not only for esports but also for potential applications of its technology in video conferencing, remote operation and telemedicine.

\section{Conclusion and future work}
This paper presents a unique but preliminary investigation of the visuo-temporal tradeoffs esports athletes make in configuring their own Dota 2 games. We find trends indicating that with a few exceptions, players prioritize temporal detail (better frame rates and latencies) over visual detail (better visual effects). In particular, most of the athletes we studied turned VSYNC off to improve temporal detail, despite the visual tearing this introduces. 

Future work should seek to confirm these trends with a more representative survey sample, and generalize them by studying other games. An experimental validation that player settings actually affect player performance as anticipated would be useful. There is a great deal of confusion among both researchers and athletes in understanding the impact of the many configuration settings available in most games, and any efforts at improving clarity would be most welcome. Finally, we hope that any lessons learned in manipulating detail to improve esports performance would be applied to improve expert performance in general.


\bibliographystyle{ACM-Reference-Format}
\bibliography{references}


\begin{thebibliography}{15}


\ifx \showCODEN    \undefined \def \showCODEN     #1{\unskip}     \fi
\ifx \showDOI      \undefined \def \showDOI       #1{#1}\fi
\ifx \showISBNx    \undefined \def \showISBNx     #1{\unskip}     \fi
\ifx \showISBNxiii \undefined \def \showISBNxiii  #1{\unskip}     \fi
\ifx \showISSN     \undefined \def \showISSN      #1{\unskip}     \fi
\ifx \showLCCN     \undefined \def \showLCCN      #1{\unskip}     \fi
\ifx \shownote     \undefined \def \shownote      #1{#1}          \fi
\ifx \showarticletitle \undefined \def \showarticletitle #1{#1}   \fi
\ifx \showURL      \undefined \def \showURL       {\relax}        \fi
\providecommand\bibfield[2]{#2}
\providecommand\bibinfo[2]{#2}
\providecommand\natexlab[1]{#1}
\providecommand\showeprint[2][]{arXiv:#2}

\bibitem[\protect\citeauthoryear{Carto}{Carto}{2021}]%
        {dota2countries}
\bibfield{author}{\bibinfo{person}{Carto}.} \bibinfo{year}{2021}\natexlab{}.
\newblock \bibinfo{booktitle}{\emph{Number of Dota 2 players per country}}.
\newblock
\urldef\tempurl%
\url{hhttps://kassgrain.carto.com/viz/c5380a68-2bc2-11e6-b452-0ecfd53eb7d3/public_map}
\showURL{%
\tempurl}


\bibitem[\protect\citeauthoryear{Chen and Thropp}{Chen and Thropp}{2007}]%
        {chen2007review}
\bibfield{author}{\bibinfo{person}{Jessie~YC Chen} {and}
  \bibinfo{person}{Jennifer~E Thropp}.} \bibinfo{year}{2007}\natexlab{}.
\newblock \showarticletitle{Review of low frame rate effects on human
  performance}.
\newblock \bibinfo{journal}{\emph{IEEE Transactions on Systems, Man, and
  Cybernetics-Part A: Systems and Humans}} \bibinfo{volume}{37},
  \bibinfo{number}{6} (\bibinfo{year}{2007}), \bibinfo{pages}{1063--1076}.
\newblock


\bibitem[\protect\citeauthoryear{Claypool and Claypool}{Claypool and
  Claypool}{2009}]%
        {claypool2009perspectives}
\bibfield{author}{\bibinfo{person}{Mark Claypool} {and} \bibinfo{person}{Kajal
  Claypool}.} \bibinfo{year}{2009}\natexlab{}.
\newblock \showarticletitle{Perspectives, frame rates and resolutions: it's all
  in the game}. In \bibinfo{booktitle}{\emph{Proceedings of the 4th
  International Conference on Foundations of Digital Games}}.
  \bibinfo{pages}{42--49}.
\newblock


\bibitem[\protect\citeauthoryear{Claypool, Claypool, and Damaa}{Claypool
  et~al\mbox{.}}{2006}]%
        {claypool}
\bibfield{author}{\bibinfo{person}{Mark Claypool}, \bibinfo{person}{Kajal
  Claypool}, {and} \bibinfo{person}{Feissal Damaa}.}
  \bibinfo{year}{2006}\natexlab{}.
\newblock \showarticletitle{The Effects of Frame Rate and Resolution on Users
  Playing First Person Shooter Games}.
\newblock \bibinfo{journal}{\emph{Proceedings of SPIE - The International
  Society for Optical Engineering}}  \bibinfo{volume}{6071} (\bibinfo{date}{01}
  \bibinfo{year}{2006}).
\newblock
\urldef\tempurl%
\url{https://doi.org/10.1117/12.648609}
\showDOI{\tempurl}


\bibitem[\protect\citeauthoryear{Janzen and Teather}{Janzen and
  Teather}{2014}]%
        {10.1145/2559206.2581214}
\bibfield{author}{\bibinfo{person}{Benjamin~F. Janzen} {and}
  \bibinfo{person}{Robert~J. Teather}.} \bibinfo{year}{2014}\natexlab{}.
\newblock \showarticletitle{Is 60 FPS Better than 30? The Impact of Frame Rate
  and Latency on Moving Target Selection}. In \bibinfo{booktitle}{\emph{CHI '14
  Extended Abstracts on Human Factors in Computing Systems}} (Toronto, Ontario,
  Canada) \emph{(\bibinfo{series}{CHI EA '14})}.
  \bibinfo{publisher}{Association for Computing Machinery},
  \bibinfo{address}{New York, NY, USA}, \bibinfo{pages}{1477–1482}.
\newblock
\showISBNx{9781450324748}
\urldef\tempurl%
\url{https://doi.org/10.1145/2559206.2581214}
\showDOI{\tempurl}


\bibitem[\protect\citeauthoryear{Kim, Spjut, McGuire, Majercik, Boudaoud,
  Albert, and Luebke}{Kim et~al\mbox{.}}{2019}]%
        {kim2019esports}
\bibfield{author}{\bibinfo{person}{Joohwan Kim}, \bibinfo{person}{Josef Spjut},
  \bibinfo{person}{Morgan McGuire}, \bibinfo{person}{Alexander Majercik},
  \bibinfo{person}{Ben Boudaoud}, \bibinfo{person}{Rachel Albert}, {and}
  \bibinfo{person}{David Luebke}.} \bibinfo{year}{2019}\natexlab{}.
\newblock \showarticletitle{Esports Arms Race: Latency and Refresh Rate for
  Competitive Gaming Tasks}.
\newblock \bibinfo{journal}{\emph{Journal of Vision}} \bibinfo{volume}{19},
  \bibinfo{number}{10} (\bibinfo{year}{2019}), \bibinfo{pages}{218c}.
\newblock


\bibitem[\protect\citeauthoryear{Liu, Claypool, Kuwahara, Sherman, and
  Scovell}{Liu et~al\mbox{.}}{2021}]%
        {10.1145/3411764.3445245}
\bibfield{author}{\bibinfo{person}{Shengmei Liu}, \bibinfo{person}{Mark
  Claypool}, \bibinfo{person}{Atsuo Kuwahara}, \bibinfo{person}{Jamie Sherman},
  {and} \bibinfo{person}{James~J Scovell}.} \bibinfo{year}{2021}\natexlab{}.
\newblock \bibinfo{booktitle}{\emph{Lower is Better? The Effects of Local
  Latencies on Competitive First-Person Shooter Game Players}}.
\newblock \bibinfo{publisher}{Association for Computing Machinery},
  \bibinfo{address}{New York, NY, USA}.
\newblock
\showISBNx{9781450380966}
\urldef\tempurl%
\url{https://doi.org/10.1145/3411764.3445245}
\showURL{%
\tempurl}


\bibitem[\protect\citeauthoryear{Musabirov, Bulygin, Okopny, and
  Sirotkin}{Musabirov et~al\mbox{.}}{2017}]%
        {10.1145/3025453.3025893}
\bibfield{author}{\bibinfo{person}{Ilya Musabirov}, \bibinfo{person}{Denis
  Bulygin}, \bibinfo{person}{Paul Okopny}, {and} \bibinfo{person}{Alexander
  Sirotkin}.} \bibinfo{year}{2017}\natexlab{}.
\newblock \showarticletitle{Deconstructing Cosmetic Virtual Goods Experiences
  in Dota 2}. In \bibinfo{booktitle}{\emph{Proceedings of the 2017 CHI
  Conference on Human Factors in Computing Systems}} (Denver, Colorado, USA)
  \emph{(\bibinfo{series}{CHI '17})}. \bibinfo{publisher}{Association for
  Computing Machinery}, \bibinfo{address}{New York, NY, USA},
  \bibinfo{pages}{2054–2058}.
\newblock
\showISBNx{9781450346559}
\urldef\tempurl%
\url{https://doi.org/10.1145/3025453.3025893}
\showDOI{\tempurl}


\bibitem[\protect\citeauthoryear{pcgamehaven.com}{pcgamehaven.com}{2020}]%
        {pcgamehaven}
\bibfield{author}{\bibinfo{person}{pcgamehaven.com}.}
  \bibinfo{year}{2020}\natexlab{}.
\newblock \bibinfo{booktitle}{\emph{What is VSync? Should you disable it, or
  enable it?}}
\newblock
\urldef\tempurl%
\url{https://pcgamehaven.com/what-is-vsync-should-you-disable-it-or-enable-it/}
\showURL{%
\tempurl}


\bibitem[\protect\citeauthoryear{Reddit.com}{Reddit.com}{2013}]%
        {redditpost}
\bibfield{author}{\bibinfo{person}{Reddit.com}.}
  \bibinfo{year}{2013}\natexlab{}.
\newblock \bibinfo{booktitle}{\emph{Is there a way to increase graphical
  quality through the config?}}
\newblock
\urldef\tempurl%
\url{https://www.reddit.com/r/DotA2/comments/15tyoo/is_there_a_way_to_increase_graphical_quality/c7puu5s/}
\showURL{%
\tempurl}


\bibitem[\protect\citeauthoryear{reddit.com}{reddit.com}{2021}]%
        {rdota2}
\bibfield{author}{\bibinfo{person}{reddit.com}.}
  \bibinfo{year}{2021}\natexlab{}.
\newblock \bibinfo{booktitle}{\emph{Dota 2 subreddit}}.
\newblock
\urldef\tempurl%
\url{https://www.reddit.com/r/DotA2}
\showURL{%
\tempurl}


\bibitem[\protect\citeauthoryear{Spjut, Boudaoud, Binaee, Kim, Majercik,
  McGuire, Luebke, and Kim}{Spjut et~al\mbox{.}}{2019}]%
        {10.1145/3355088.3365170}
\bibfield{author}{\bibinfo{person}{Josef Spjut}, \bibinfo{person}{Ben
  Boudaoud}, \bibinfo{person}{Kamran Binaee}, \bibinfo{person}{Jonghyun Kim},
  \bibinfo{person}{Alexander Majercik}, \bibinfo{person}{Morgan McGuire},
  \bibinfo{person}{David Luebke}, {and} \bibinfo{person}{Joohwan Kim}.}
  \bibinfo{year}{2019}\natexlab{}.
\newblock \showarticletitle{Latency of 30 Ms Benefits First Person Targeting
  Tasks More Than Refresh Rate Above 60 Hz}. In
  \bibinfo{booktitle}{\emph{SIGGRAPH Asia 2019 Technical Briefs}} (Brisbane,
  QLD, Australia) \emph{(\bibinfo{series}{SA '19})}.
  \bibinfo{publisher}{Association for Computing Machinery},
  \bibinfo{address}{New York, NY, USA}, \bibinfo{pages}{110–113}.
\newblock
\showISBNx{9781450369459}
\urldef\tempurl%
\url{https://doi.org/10.1145/3355088.3365170}
\showDOI{\tempurl}


\bibitem[\protect\citeauthoryear{Steamcharts.com}{Steamcharts.com}{2021}]%
        {dota2numbers}
\bibfield{author}{\bibinfo{person}{Steamcharts.com}.}
  \bibinfo{year}{2021}\natexlab{}.
\newblock \bibinfo{booktitle}{\emph{Dota 2 - Steam Charts}}.
\newblock
\urldef\tempurl%
\url{https://steamcharts.com/app/570}
\showURL{%
\tempurl}


\bibitem[\protect\citeauthoryear{Steamcommunity.com}{Steamcommunity.com}{2016}]%
        {steamcommunity}
\bibfield{author}{\bibinfo{person}{Steamcommunity.com}.}
  \bibinfo{year}{2016}\natexlab{}.
\newblock \bibinfo{booktitle}{\emph{Optimization / Tweak for DotA 2 Gaming}}.
\newblock
\urldef\tempurl%
\url{https://steamcommunity.com/sharedfiles/filedetails/?id=339319323}
\showURL{%
\tempurl}


\bibitem[\protect\citeauthoryear{Watson, Shrivastava, and Gavane}{Watson
  et~al\mbox{.}}{2019}]%
        {10.1145/3320286}
\bibfield{author}{\bibinfo{person}{Benjamin Watson}, \bibinfo{person}{Rachit
  Shrivastava}, {and} \bibinfo{person}{Ajinkya Gavane}.}
  \bibinfo{year}{2019}\natexlab{}.
\newblock \showarticletitle{The Effects of Adaptive Synchronization on
  Performance and Experience in Gameplay}.
\newblock \bibinfo{journal}{\emph{Proc. ACM Comput. Graph. Interact. Tech.}}
  \bibinfo{volume}{2}, \bibinfo{number}{1}, Article \bibinfo{articleno}{5}
  (\bibinfo{date}{June} \bibinfo{year}{2019}), \bibinfo{numpages}{13}~pages.
\newblock
\urldef\tempurl%
\url{https://doi.org/10.1145/3320286}
\showDOI{\tempurl}


\end{thebibliography}

\end{document}


\title{Better Frame Rates or Better Visuals? \\ An Early Report of Esports Player Practice in Dota 2}

\author{Arjun Madhusudan}
\affiliation{%
  \institution{North Carolina State University}
  \city{Raleigh}
  \state{North Carolina}
  \country{USA}
  }
\email{amadhus2@ncsu.edu}

\author{Benjamin Watson}
\affiliation{%
  \institution{North Carolina State University}
  \city{Raleigh}
  \state{North Carolina}
  \country{USA}
  }
\email{bwatson@ncsu.edu}

\maketitle



\appendix
\section{Settings Description}
\label{app:settings_config}
Below is the complete list of the 15 settings we studied and their explanations. We compiled these with extensive testing and online research in documentation and forums. Unfortunately, we could not locate one definitive resource that clearly defines all of these, let alone the full suite of 450 Dota 2 settings.

\begin{itemize}
    \item {{\bfseries \textit{useadvanced}}: Indicates the mode of customization used. 0 uses presets provided by the game to manage other settings listed below as a group, while 1 allows the player to customize each option individually.}
    \item {{\bfseries \textit{mat\_vsync}}: Indicates whether the player has VSYNC on or off. When VSYNC is on, it improves visual detail by adding delay to synchronize frames to refresh rate. When it is off, it instead improves temporal detail by removing this delay --- but without synchronization, the display shows parts of multiple frames, with visual ``tears'' between them.}
    \item {{\bfseries \textit{r\_texture\_stream\_mip\_bias}}: Changes texture resolution by changing the mipmap bias. 0 uses coarse textures, 1 medium, and 2 fine. We considered values 0 and 1 as reduced visuals.}
    \item {{\bfseries \textit{mat\_viewportscale}}: Scaling factor for the viewport. If less than 1, it limits visual detail by using fewer pixels in rendered imagery.}
    \item {{\bfseries \textit{cl\_globallight\_shadow\_mode}}: Level of detail in shadows, with 0 being off, 1 being blob (a roughly circular shadow under any object), 2 being stencil (shadows resembling the silhouette of the casters), and 3 being ultra (rendering shadows for additional objects such as the optional ambient creatures). We considered values 0 and 1 as limiting visual detail, while 2 and 3 did not.}
    \item {{\bfseries \textit{cl\_particle\_fallback\_base}}: Controls level of detail in special effects such as glow, particles, and smoke, with 0 being low, 2 medium and 4 high. We considered values 0 and 2 as limiting visual detail, while 4 did not.}
    \item {{\bfseries \textit{dota\_ambient\_creatures}}: Adds aesthetic environmental creatures to the game visuals. Set to 0 (off) or 1 (on).}
    \item {{\bfseries \textit{dota\_cheap\_water}}: Changes the quality of flowing water. When set to 0, water is static and does not reflect or refract anything. 1 adds reflection, refraction, flow, and splashes when players interact with it.}
    \item {{\bfseries \textit{r\_dota\_allow\_wind\_on\_trees}}: Enables dynamic foliage. 0 uses static trees and shrubs while 1 adds swaying animation to them.}
    \item {{\bfseries \textit{r\_grass\_quality}}: Controls level of detail in grass textures. 0 uses static texture maps while 1 adds depth and animation to the grass, such as moving when walking through it.}
    \item {{\bfseries \textit{dota\_ambient\_cloth}}: Enables realistic cloth simulation such as character attire, flags, and banners waving in the wind. Can be set to 0 (off) or 1 (on).}
    \item {{\bfseries \textit{r\_deferred\_additive\_pass}}: Enables deferred (pre-pass) lighting effects. Toggles between 0 (off) and 1 (on).}
    \item {{\bfseries \textit{r\_deferred\_simple\_light}}: Can toggle additional aesthetic world lighting, such as torches and lamps, between 0 (off) and 1 (on) for better visuals.}
    \item {{\bfseries \textit{r\_deferred\_specular}}: Toggles specular reflections on world objects between 0 (off) and 1 (on).}
    \item {{\bfseries \textit{r\_deferred\_specular\_bloom}}: Toggles specular reflections and bloom on characters between 0 (off) and 1 (on).}
    \item {{\bfseries \textit{r\_ssao}}: Allows the use of screen space anti-aliasing. Can be toggled between 0 (off) or 1 (on).}
\end{itemize}
%
\appendix







